\documentclass[a4paper,12pt]{article}
\usepackage{graphicx}
\usepackage{hyperref}
\usepackage{infnprep}
\usepackage[top=2.5cm, bottom=3cm, left=3cm, right=3cm]{geometry}
\newcommand{\Header}{
  \resizebox{15cm}{!}{
  \begin{tabular}{rl}
  \includegraphics[width=5cm, trim={50 100 0 0}]{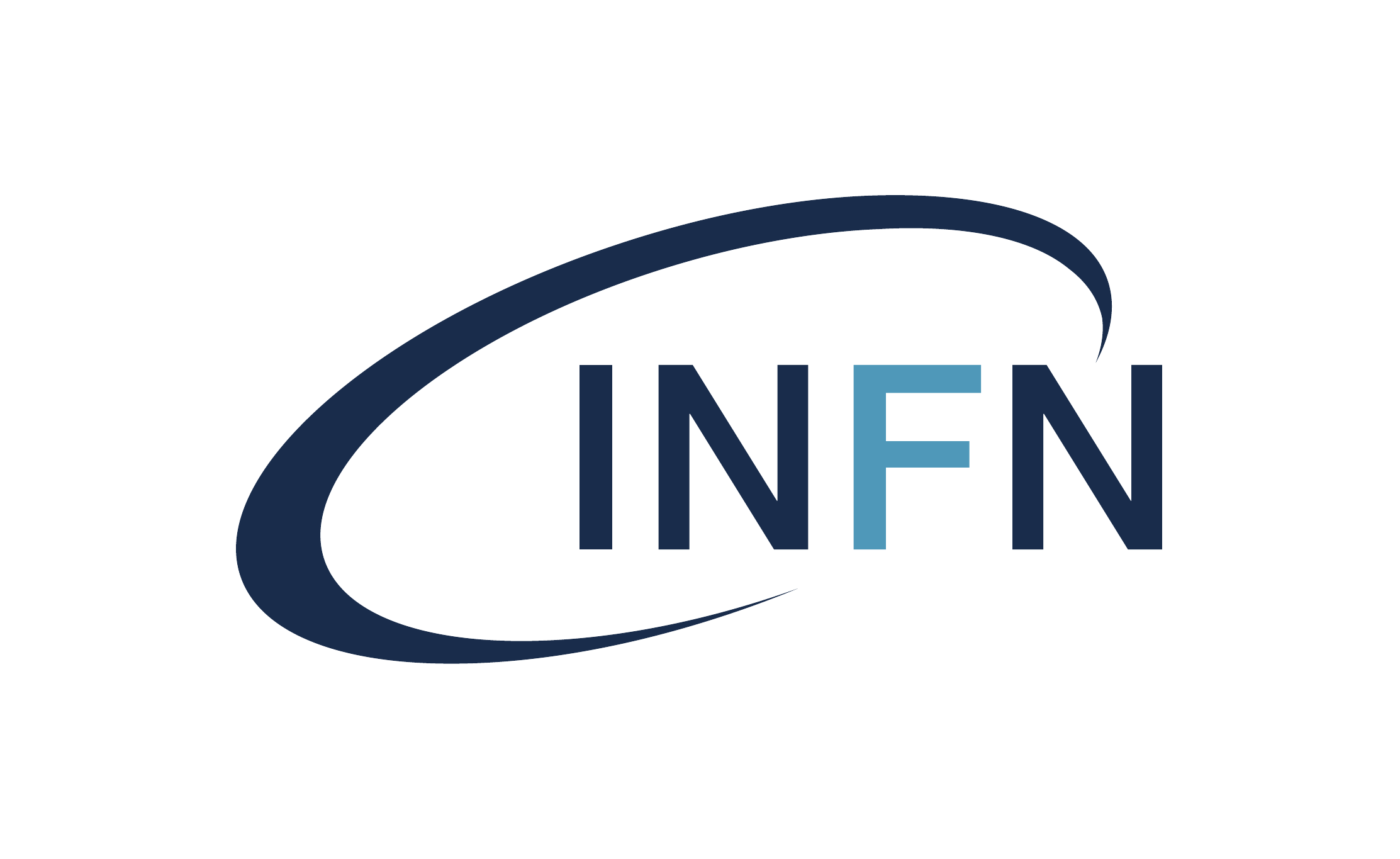} & {\LARGE\sffamily ISTITUTO NAZIONALE DI FISICA NUCLEARE}\\
      \\
  \end{tabular}
  }
\begin{center}
      {\large\sffamily Sezione di Roma}\\
\end{center}
    \renewcommand{\arraystretch}{1}
\vskip 0.5cm
\rule{15.0cm}{0.09mm}
\vskip 1.5cm
  \begin{flushright}
      {\underline{\bf INFN-19-20/ROMA1}}\\    
      {\small\bf 30 Novembre 2019} \\      
  \end{flushright}
}
\begin{document}
\begin{titlepage}
\title
  {\Header \large \bf Ideas for extending the Frascati LINAC positron beam pulses for the resonant search of a $X$(17 MeV) boson.}
\author{
   Paolo Valente\\
{\it INFN, Sezione di Roma, P.le Aldo Moro 2, I-00185 Roma, Italy} 
} 
\maketitle
\baselineskip=14pt
\begin{abstract}
The results on the so-called $^8$Be anomaly, recently corroborated by similar experimental evidence in the radiative transitions of excited $^4$He nuclei, could be justified by the creation of a new particle with a mass of $m_X\simeq16.7$~MeV/$c^2$. 

The PADME experiment, designed for searching light dark sector particles, like a dark photon or an axion-like particle, both in $\gamma +$ missing energy and $e^+ e^-$ final states, has the potential of performing a completely independent search, also exploiting the cross-section enhancement at the resonance $\sqrt{s}\simeq m_X$. In the case of the $X$(17 MeV) boson, this corresponds to a positron energy of 282 MeV when annihilating on electrons at rest.

In order to keep the pile-up and the over-veto probabilities under control, the positron beam hitting the PADME active target should be as much diluted in time as possible. PADME has already collected a first data-set at the Frascati beam-test facility, using the positron beam accelerated by the DA$\Phi$NE LINAC, with maximum length of $\sim$200~ns and energy 490--550 MeV.

In this note, the possible modifications to the RF system of the LINAC, aiming at further extending the pulse length at the expenses of the maximum beam energy, are briefly discussed.

\end{abstract}
\vspace*{\stretch{2}}
\begin{flushleft}
\end{flushleft}
\begin{flushright}
  \vskip 3cm
\small\it Published by \\
Laboratori Nazionali di Frascati
\end{flushright}
\end{titlepage}
\pagestyle{plain}
\setcounter{page}2
\baselineskip=17pt
\section{Introduction}
\subsection{Dark sector searches and the PADME experiment}
The search for light particles in the MeV-GeV range from a dark sector has attracted an increasing attention, also due to the lack of positive observation of WIMP or other heavy dark matter candidates. 

A rich experimental program has flourished all around the world in the recent years, as well as phenomenological studies~\cite{ref1}, also in connection with other anomalies. In particular contributions from new particles can justify the tension between the measured and calculated muon $g-2$~\cite{ref2}, and a new, light boson has been advocated for justifying the unexpected bump in the angular distribution of internally created $e^+  e^-$ pairs of $^8$Be nuclei, only for transitions above $\simeq$17~MeV (in particular in the 18.15~MeV nuclear M1 transitions with $J^\pi=1^+\to0^+$)~\cite{ref3}.
 
The latter result, indicating the possible creation of a new particle with a mass of $m_X\simeq16.7$~MeV/$c^2$, has been recently reinforced by the observation~\cite{ref4}, by the same ATOMKI group, of a similar bump in the 21.01 MeV M0 transitions ($J^\pi=0^-\to0^+$) of $^4$He nuclei, yielding practically the same mass for the hypothetical intermediate particle, even though the electron-positron angular distribution exhibits a peak at a different angle.

At the INFN Frascati laboratories (LNF) is currently installed the PADME experiment~\cite{ref5}, on the first beamline of the beam-test facility (BTF-1)~\cite{ref6}, which makes use of the DA$\Phi$NE LINAC. PADME aims at searching in the most general way dark mediators ($X$) in the annihilation of a 550~MeV positron beam onto the electrons of a thin target (0.1 mm active diamond), looking for a peak in the missing mass spectrum of $e^+  e^-\to\gamma+X$, thanks to a fine-grained, high-resolution BGO crystal calorimeter, a fast, small-angle PbF$_2$ calorimeter, a set of scintillating bars veto detectors in a large vacuum vessel. 

This approach allows to search for the so-called ``invisible'' decays of the dark photon, i.e. when the mediator is heavy enough to decay dominantly into pairs of lighter particles in the dark sector: 
$A’\to\chi\overline{\chi}$, with $m_{A^\prime}>2m_\chi$. The maximum positron beam energy of 550~MeV sets the maximum mass $m_{A^\prime}$  that can be created in the fixed-target annihilations to $\simeq$24~MeV/$c^2$.

\subsection{PADME positron beam requirements}
One of the main requirements for PADME is to dilute in time the positron beam pulse as much as possible, in order to reduce the event pile-up in the detectors and the over-veto probability. The maximum achievable luminosity is thus basically limited by the longest achievable high-energy positron beam pulse. 

A first significant data sample was collected in a run from November 2018 to February 2019, using beam pulses with intensity in the range $0.5$ to $3\times10^4$ positrons, and duration up to 200~ns, mainly using a ``secondary'' beam (see next section). This data set is currently being analyzed, and a new physics run, using mainly a ``primary'' positron beam, is planned in Spring 2020~\cite{ref7}.

Concerning the positron energy, for dark photon or generic light dark sector particles the requirement is to have the highest possible beam energy $E_+$, since it sets the maximum mass value that can be produced: $M=\sqrt{2m_e E_+}$. In original PADME proposal\cite{ref5} $E_+=550$~MeV was assumed, just considering the maximum energy of the positron beam produced by the DA$\Phi$NE LINAC.

It has been noticed~\cite{ref8} that in searching light dark mediators, such as a kinetically mixed dark photon $A’$ or the $X$(17~MeV) boson, the production rate in $e^+ e^-$ annihilations is greatly enhanced by running close to the resonance $\sqrt{s}\simeq m_{A^\prime}$.
This has driven the idea of running the PADME experiment tuning the positron energy close to the resonance $\sim$17~MeV/$c^2$, i.e. setting the beam to an energy $E_+\sim$282~MeV, and looking for the ``visible'' decay $X\to e^+ e^-$, in order to confirm the boson indicated by the ATOMKI $^8$Be anomaly in a completely independent way in the direct production $e^+ e^-\to X$.

In order to allow scanning across the resonance (and have a minimum margin), the beam energy has to be approximately 300~MeV, so that the Frascati LINAC turns to be one of the very few accelerator facilities providing sufficiently high-energy positrons~\cite{ref9}. 

It has been also noticed\cite{ref10} that an additional bonus would come from the use of a thick target in searching for dark photon production in fixed-target positron on electron annihilations, thanks to the enhancement of the production rate: the high number of secondary positrons in the electromagnetic shower allows in fact producing the dark particle both in the non-resonant $e^+ e^-\to A^\prime\gamma$ and resonant $e^+ e^-\to A^\prime$ reactions; this also applies to the $X$ boson case.
 
\begin{figure}[htbp]
\centering
\includegraphics[width=0.7\textwidth]{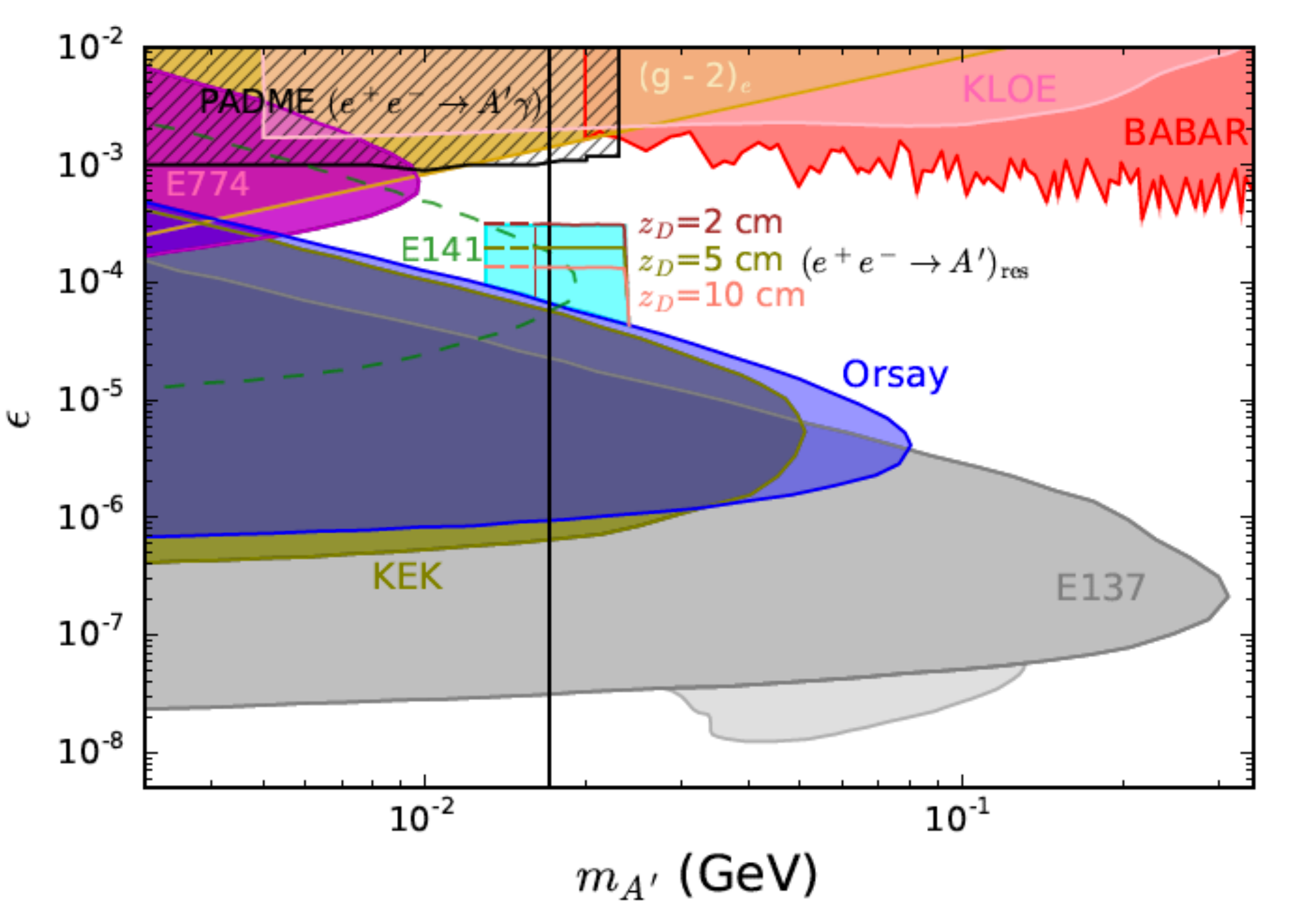}
 \caption{
      Sensitivity for dark photon $A^\prime$ in the plane mass $m_{A^\prime}$ vs. kinetic mixing parameter $\epsilon$; the vertical line is drawn at 17 MeV/$c^2$, the cyan-shaded areas show the exclusion region for PADME in thick-target mode for different thickness values (reproduced from Ref.~\cite{ref8}).
    \label{fig1} }
\end{figure}

The choice of the thickness of the target affects the range of accessible lifetime of the hypothetical boson, translating into a different sensitivity in the coupling (as shown in Fig.~\ref{fig1}, reproduced from Ref.~\cite{ref8}, but it also determines the fraction of the electromagnetic shower surviving the target and thus entering the detector. The acceptable background level would then set the limit on the beam intensity and hence the achievable luminosity.

This case is very different from the thin target, invisible decays search of a dark photon, in the “standard” PADME experiment: on one hand the target greatly multiplies the number of secondary particles, unless its thickness allows an almost full containment of the shower ($>20 x_0$), on the other, the spectrum of the surviving particles will be spread out down to very low energy, while the genuine pairs coming from the $X$ decays have to reconstruct a $\sim$17~MeV invariant mass. 

In any case, a high-intensity positron beam is desirable, up to the maximum achievable charge from the DA$\Phi$NE LINAC. 
Again, thanks to the fact that a constant current can be extracted from the thermo-ionic gun, increasing the accelerated pulse length would also increase the beam charge. 
\section{Long positron pulses with SLED compression}
\subsection{DA$\Phi$NE LINAC}
The Frascati LINAC was originally designed as injector of the DA$\Phi$NE collider (see for instance Ref.~\cite{ref15}), running at the $\Phi$ meson resonance (1020~MeV), i.e. for producing high-current electron and positron pulses at 510 MeV. It is a $\sim$60~m long, S-band (2856 MHz) linear accelerator, made up by 15 travelling-wave, 2/3$\pi$, SLAC-type, 3~m long (84 cells), constant gradient accelerating sections, fed by four 45~MW klystrons (Thales TH2128-C) with RF compression~\cite{ref14}; the general RF scheme is shown in the following Fig.~\ref{fig2}.
\begin{figure}[htbp]
\centering
\includegraphics[width=\textwidth]{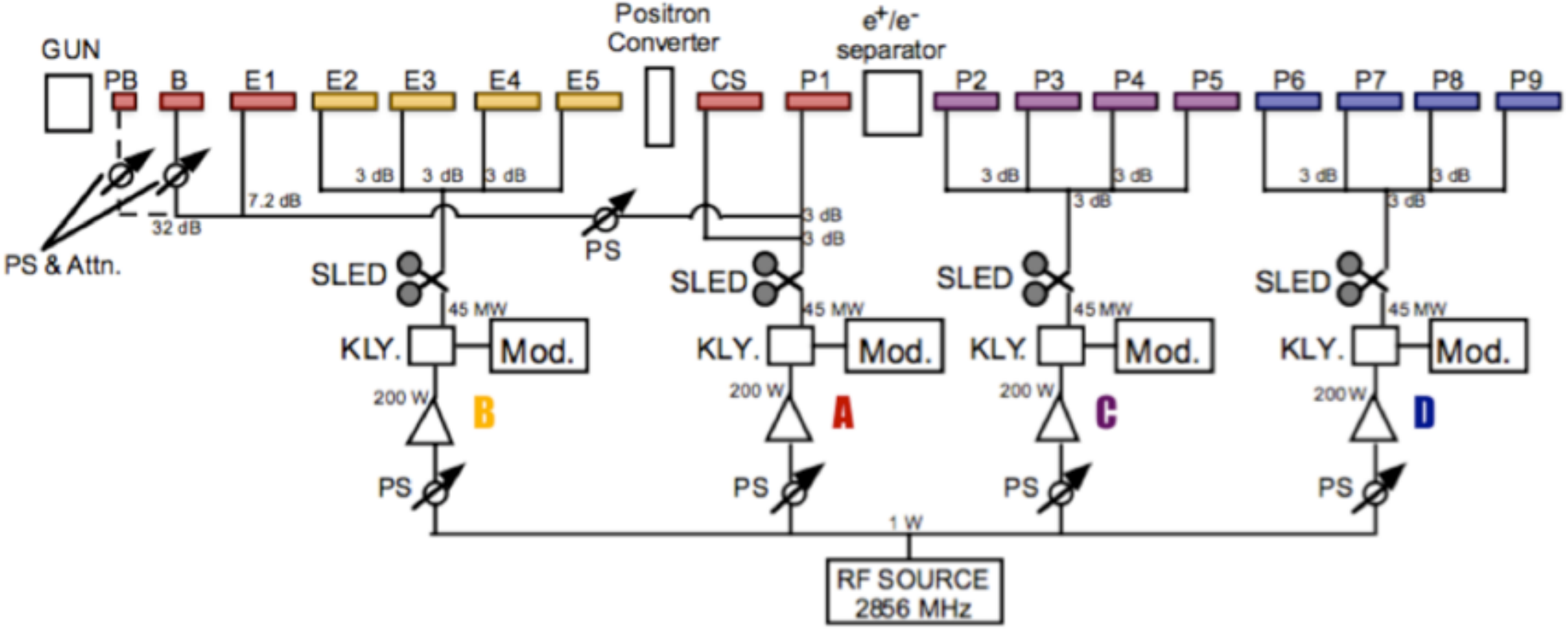}
 \caption{
      General RF scheme of the DA$\Phi$NE LINAC.
    \label{fig2} }
\end{figure}

The beam length entering the accelerating cavities is determined by the duration of the electron pulse extracted from the LINAC gun~\cite{ref11}. The thermo-ionic, triode gun, allows extracting an electron flow from the cathode of several A, by applying a rectangular waveform of adjustable length in the range 1.5~ns to 5~$\mu$s. The gun deck is operated at $\sim$100~kV potential, so that two structures (a single cell ``pre-buncher'' and a five-cell ``buncher'') have to be used for further increasing the electron velocity, up to 0.75~$c$, packing them in the RF buckets.

RF compression was introduced in order to get a higher gradient with fewer klystrons, and this was possible since the LINAC beam pulses have to be short. In particular electron and positron macro-bunches have to be $<13$~ns in order to be injected into the intermediate damping ring (RF=74~MHz, i.e. 1/5 of the main rings). The long RF pulse produced by the klystrons is compressed by means of the so-called SLED device (SLAC Energy Doubler), in which a higher peak power is achieved at expenses of the pulse length, yielding a higher accelerating voltage while leaving unchanged the average power. 

\begin{figure}[htbp]
\centering
\includegraphics[width=0.7\textwidth]{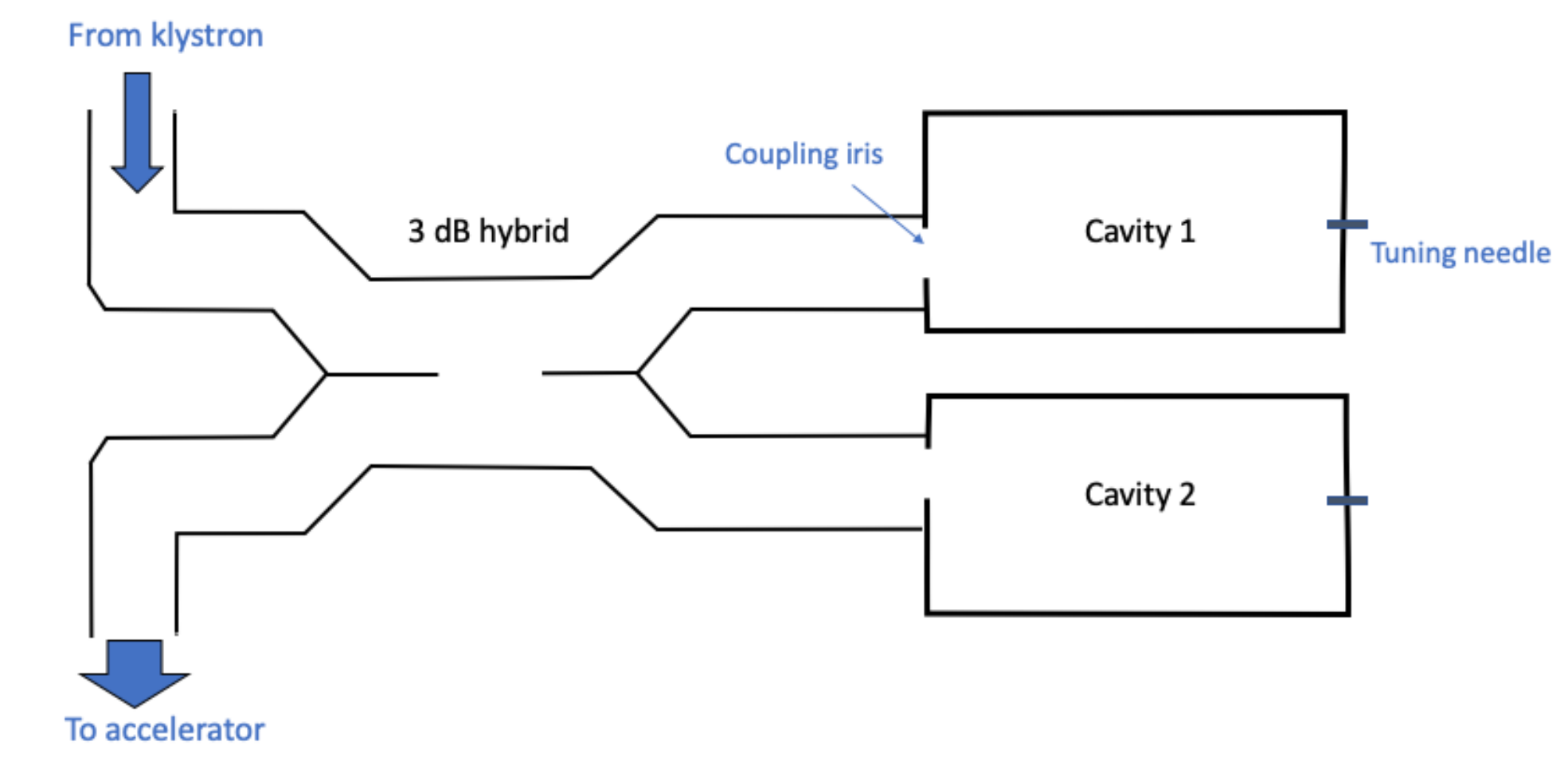}
 \caption{
      SLAC Energy Doubler (SLED) scheme.
    \label{fig3} }
\end{figure}

The SLED concept was developed at SLAC in the 1970’s~\cite{ref12}: the flat RF power from the klystron is first stored in two identical, high quality factor, cylindrical cavities in TE$_{015}$ mode, coupled by means of a 3~dB hybrid coupler. The RF energy stored for a large fraction of the pulse is discharged in a short time towards the accelerator; by reversing of 180$^\circ$ the phase of the input RF the emitted field from the cavities and the reflected field added at the output port of 3~dB coupler combine into the sharply-peaked high-power pulse. The conceptual scheme of the SLED is shown in Fig.~\ref{fig3}.

\begin{figure}[htbp]
\centering
\includegraphics[width=0.7\textwidth]{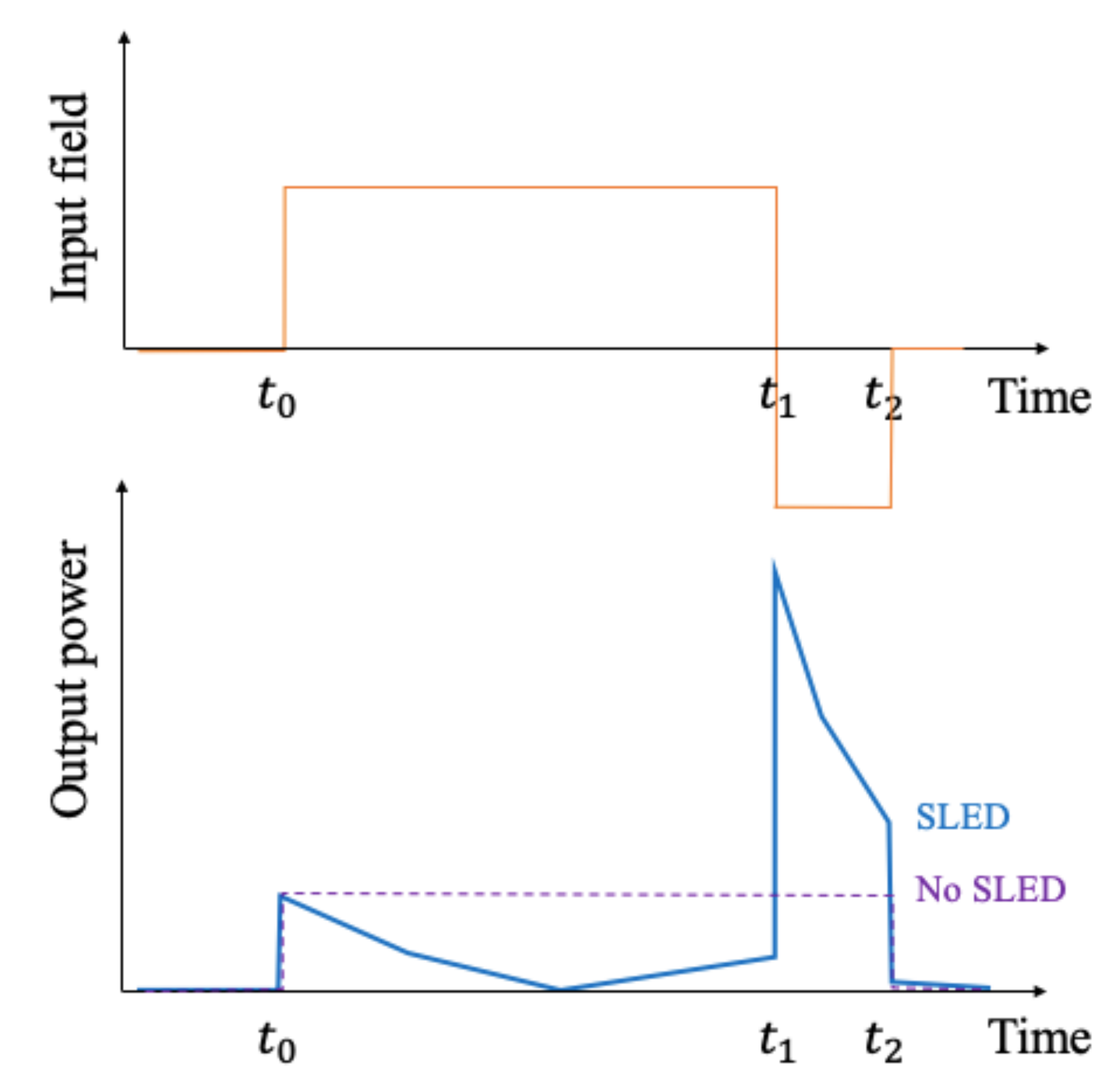}
 \caption{
      Scheme of the RF compression (adapted from Ref.~\cite{ref12}): the flat shape between the start ($t_0$) and the end ($t_2$) of the RF power from the klystron (bottom, purple), is transformed at the output of the SLED (bottom, blue) in a shape with a much higher amplitude peaking at the time of phase inversion $t_1$ (top, orange).
    \label{fig4} }
\end{figure}

The cavities are over-coupled in order to get the highest possible gain, in the case of the DA$\Phi$NE LINAC, the unloaded quality factor of the cavities is $Q_0\sim$ $10^5$, and the coupling is $\beta\sim5$. These values give an external and loaded quality factors of $Q_e=20000$ and $Q_L=16667$ respectively, a filling time of $t_f=2Q_L/\omega\sim1.9 \mu$s and a bandwidth of $\sim170$~kHz.
The shape of the input and compressed RF power and of the timing are shown in Fig.~\ref{fig4}: the klystron wave is switched on at time $t_0$, the phase inversion is applied at $t_1$, approximately $\sim$800~ns before the end of the pulse $t_2$ (4.5~$\mu$s after the $t_0$), in order to allow proper filling of the accelerating sections. The output power is enhanced by a factor $\sim3$, so that the energy gain is increased by a factor $\sim1.7$.

\begin{figure}[htbp]
\centering
\includegraphics[width=0.7\textwidth]{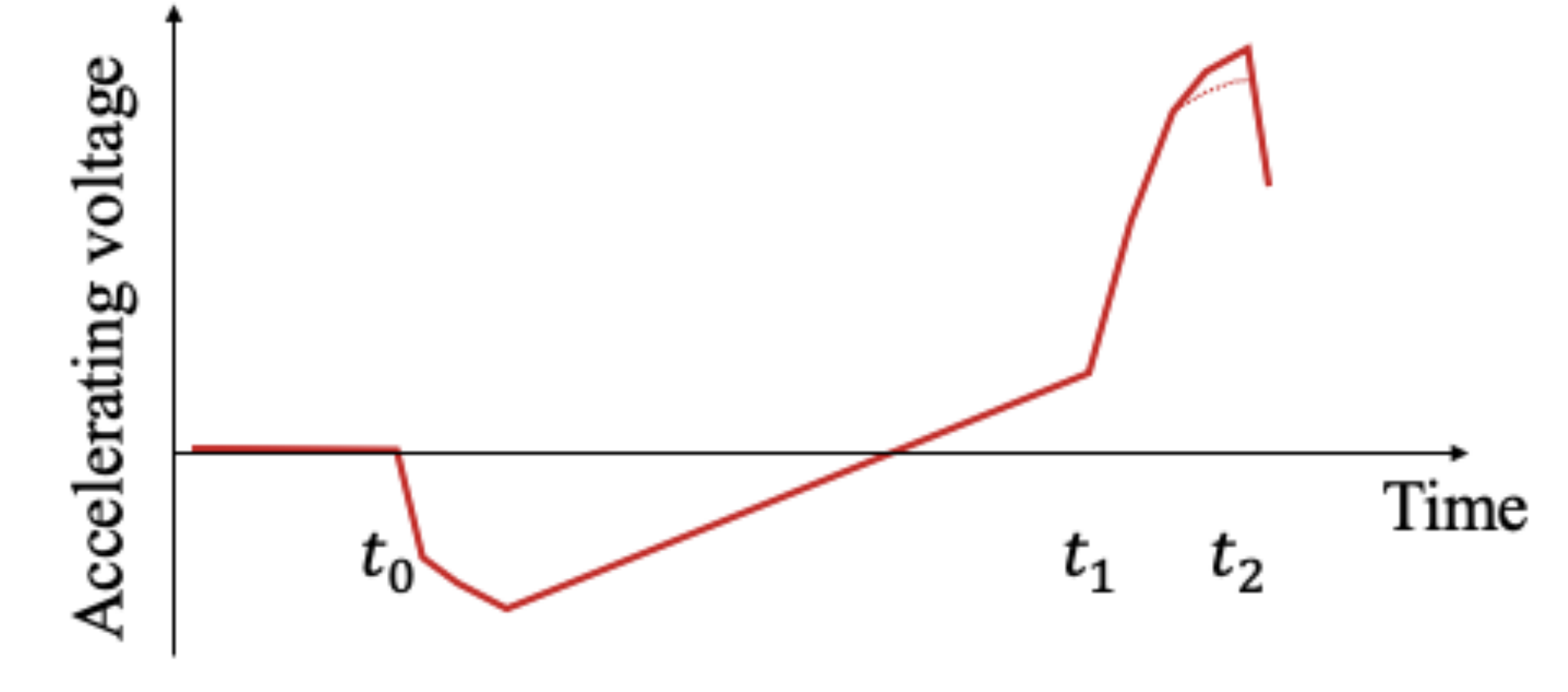}
 \caption{
      Accelerating voltage as result of the SLED compression (adapted from Ref.~\cite{ref12,ref13}).
    \label{fig5} }
\end{figure}

The resulting voltage in the accelerating structures is shown in Fig.~\ref{fig5}: it has a peak at the $t_2$ and then a sharp falling edge. While this is perfectly adequate for $\sim$10~ns beam pulses, such a shape makes difficult accelerating longer pulses, even though the peak gets smoother as the beam charge increases due to the longer pulses due to the beam-loading effects~\cite{ref13}. 

\subsection{PADME Run-1 conditions}
\label{run1}.
One of the main requirements for PADME is to dilute in time the positron beam pulse as much as possible, in order to reduce the event pile-up in the detectors and the over-veto probability. The maximum beam intensity during the first PADME run (Run-1, from November 2018 to March 2019) was adjusted in order to stay at a typical particle density of $n_{e^+}\simeq10^2\times \Delta t$ (in~ns). 

During the Run-1, beam pulses up to 200~ns have been routinely obtained~\cite{ref14}, essentially by tuning the delays of the modulators with respect to the gun timing and shifting away the beam pulse from the maximum voltage. Referring the shape of the accelerating voltage shown in Fig.~\ref{fig4}, this is possible since the macro-bunch to be accelerated extends along the rising slope of the voltage before the drop at the end of the RF pulse at $t_2$. This is achieved by advancing the time of the beam pulse, i.e. the gun reference time, with respect to the RF. The first particles in the macro-bunch will feel a lower accelerating voltage, resulting in a lower final energy of the pulse head with respect to the tail. 

\begin{figure}[htbp]
\centering
\includegraphics[width=0.65\textwidth]{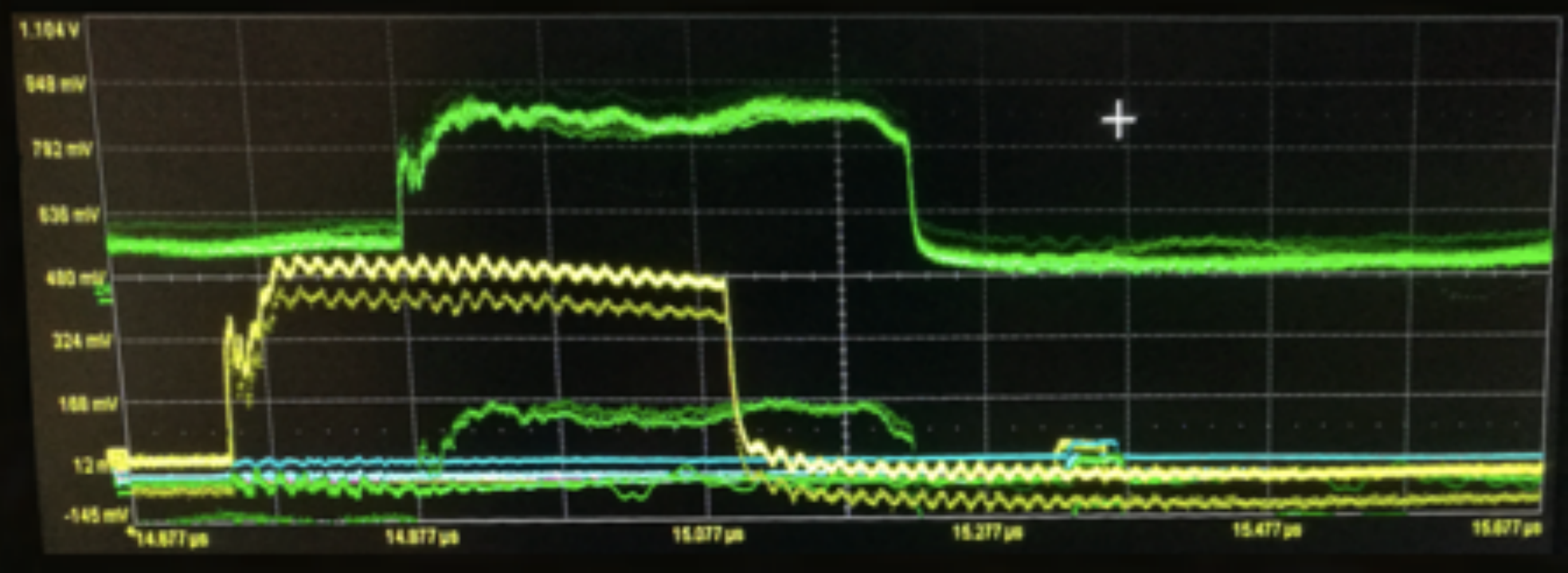}
 \caption{
      Beam pulses at the electron gun (yellow trace) and at the end of the LINAC (green trace), as measured by two of the LINAC integrating current toroids, with optimal timing of the beam pulse with respect to the RF.
    \label{fig6} }
\end{figure}

Extending the beam pulse further towards the phase inversion time $t_1$, implies a lower and lower voltage, i.e. a larger energy spread, so that the limitation on the maximum pulse length will be given by the energy acceptance of the LINAC and of the transfer-line. As a consequence of the shape of the accelerating voltage also the density of particles inside the beam macro-pulse will depend on the time, and since the accelerating structures are fed by four independent RF stations, the final distribution inside the beam pulse will depend on the power in the four modulators, i.e. on the values of the RF power and phase, and on their relative time shifts. Optimizing all those parameters it was then possible to get a fairly flat beam pulse, as shown in the example in Fig.~\ref{fig6}.

\begin{figure}[htbp]
\centering
\includegraphics[width=0.65\textwidth]{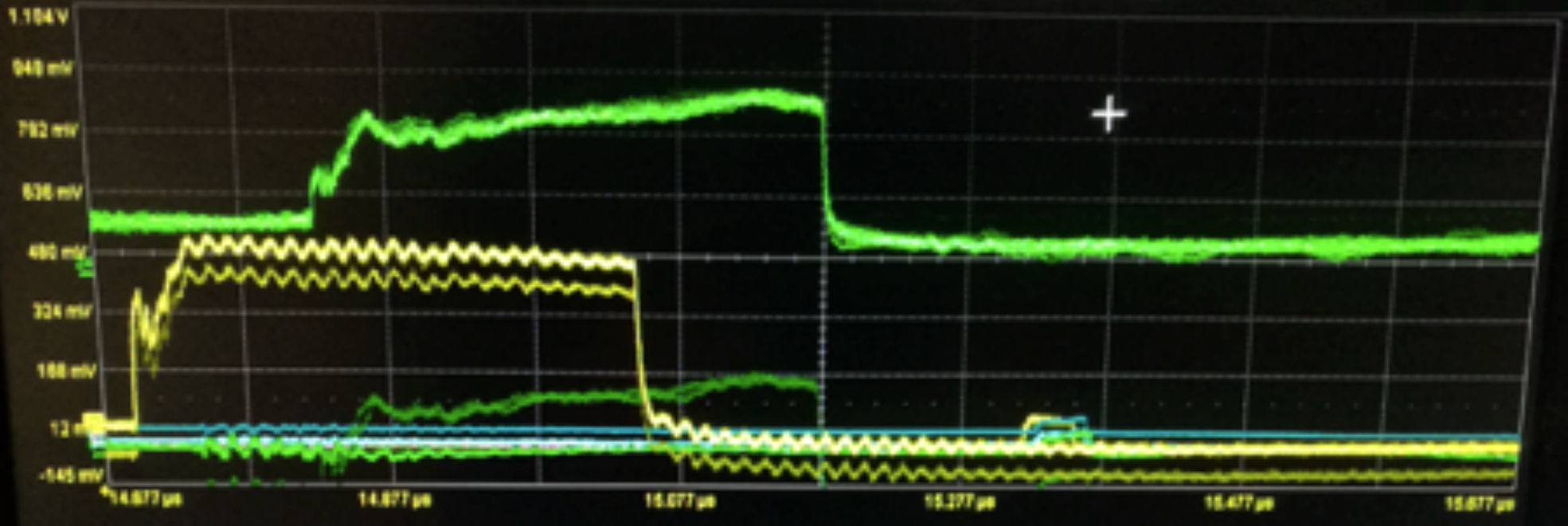}
\includegraphics[width=0.3\textwidth]{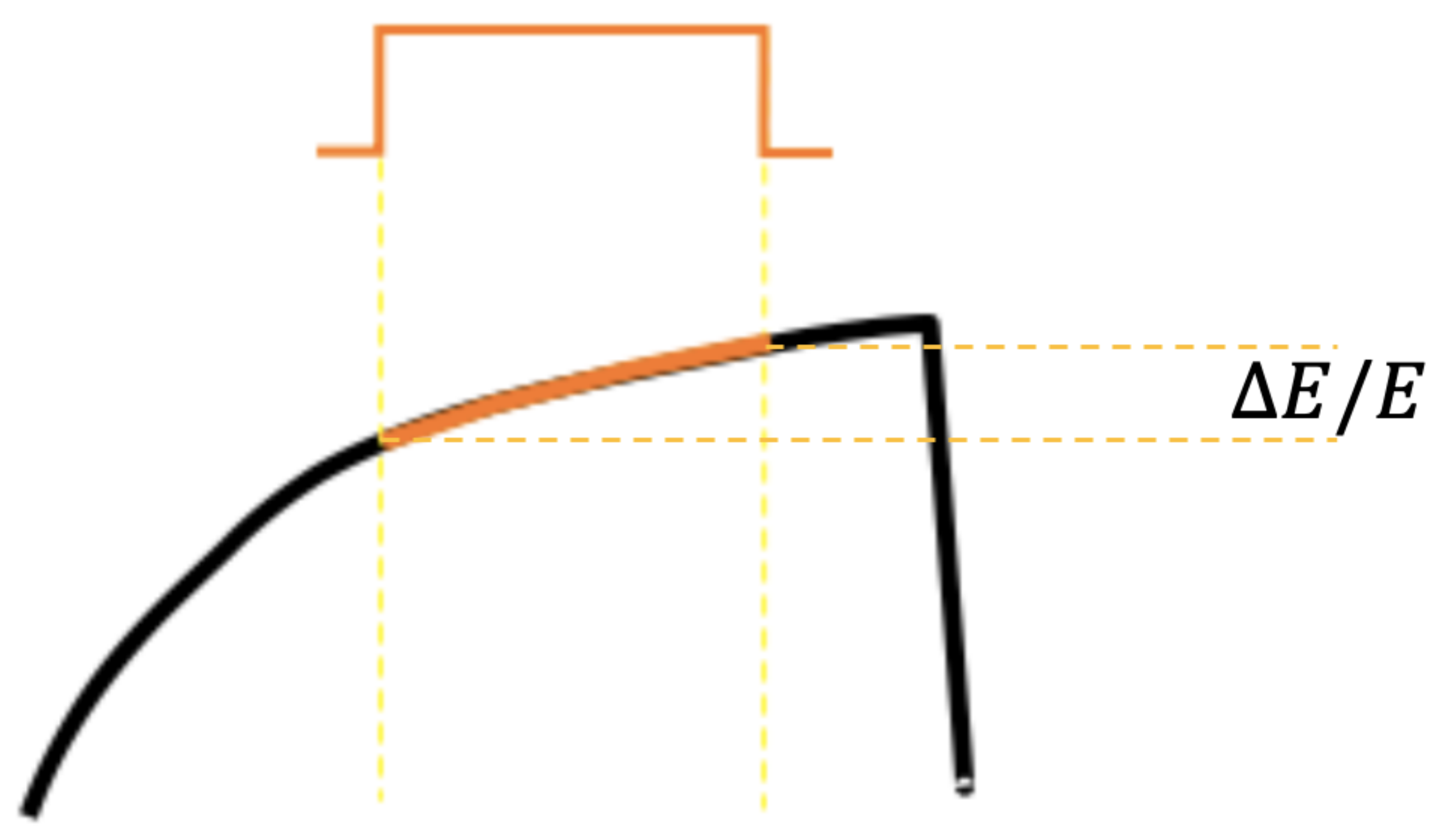}\\
\includegraphics[width=0.65\textwidth]{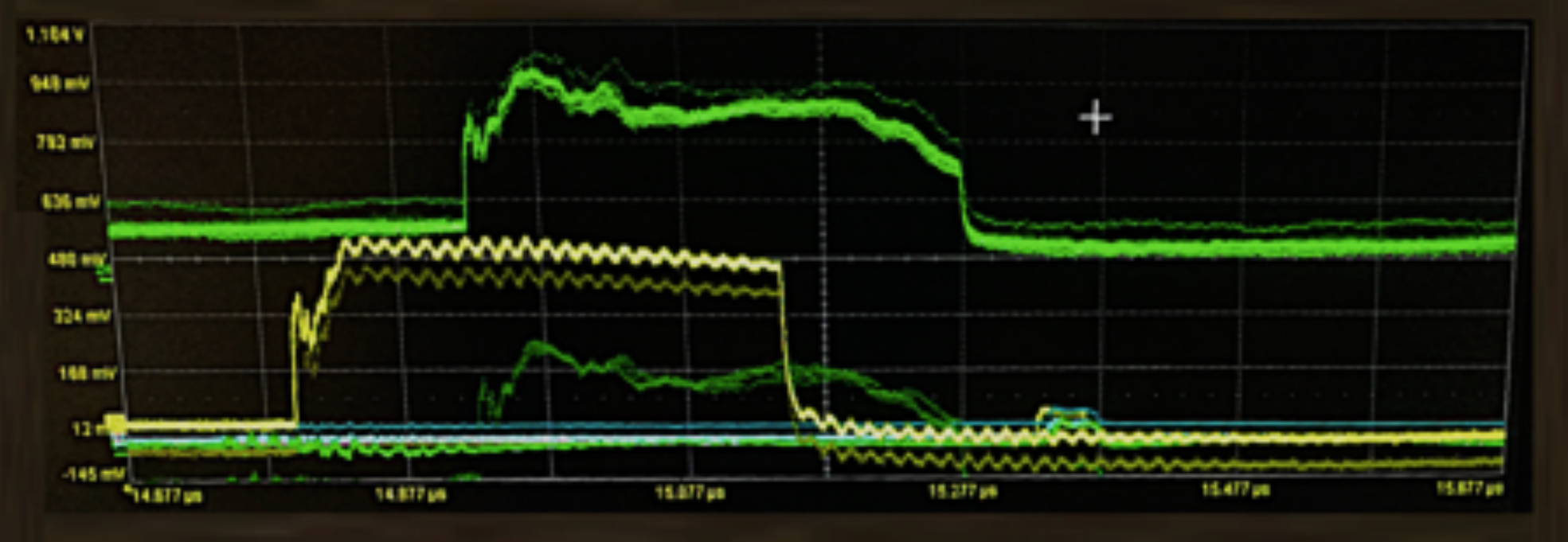}
\includegraphics[width=0.3\textwidth]{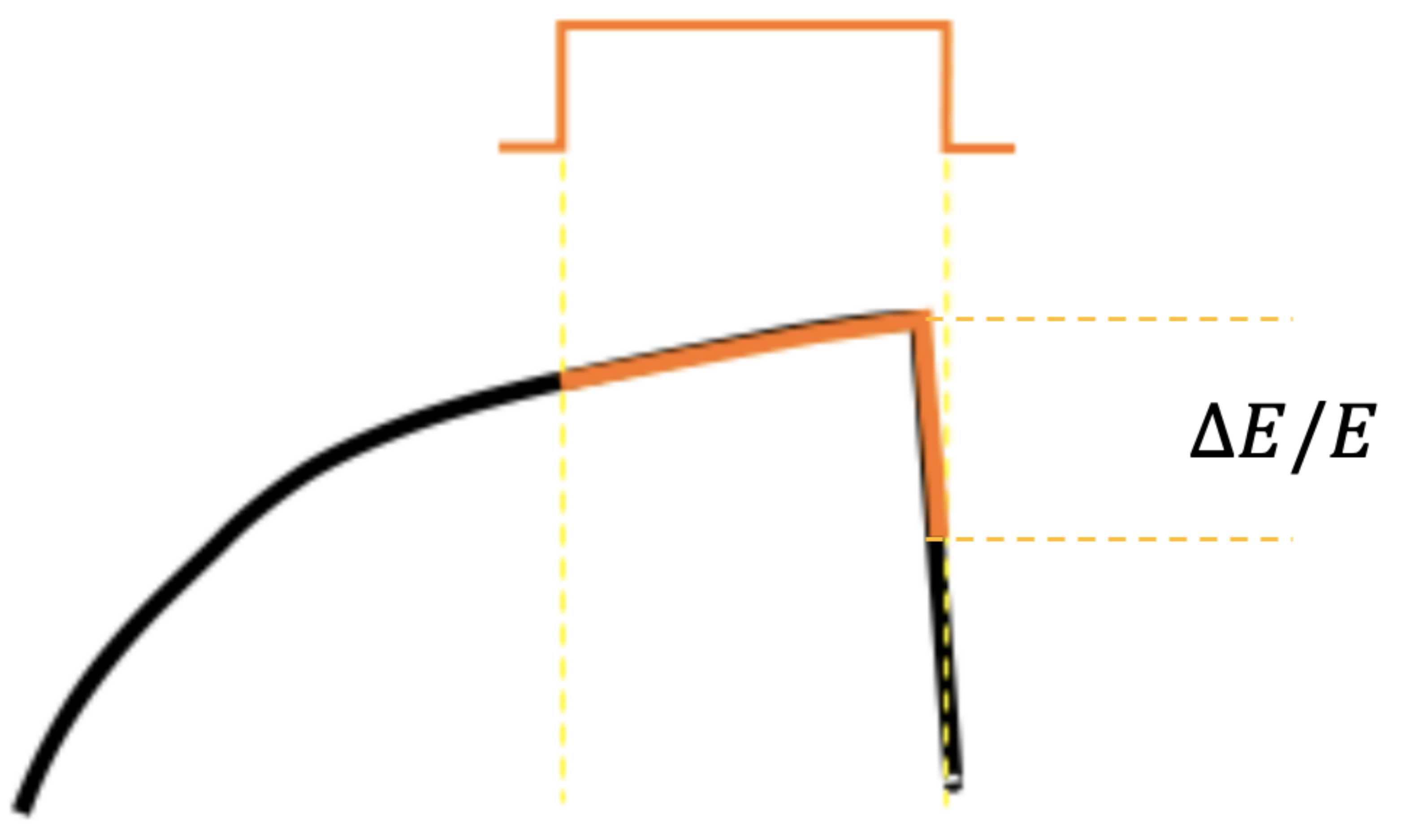}
 \caption{
      Beam pulses at the electron gun (yellow trace) and at the end of the LINAC (green trace), as measured by two of the LINAC integrating current toroids, when the gun is advanced by 50~ns (top) or delayed by 50~ns (bottom) with respect to the optimal timing with the RF.   
    \label{fig7} }
\end{figure}

\begin{figure}[htbp]
\centering
\includegraphics[width=0.5\textwidth]{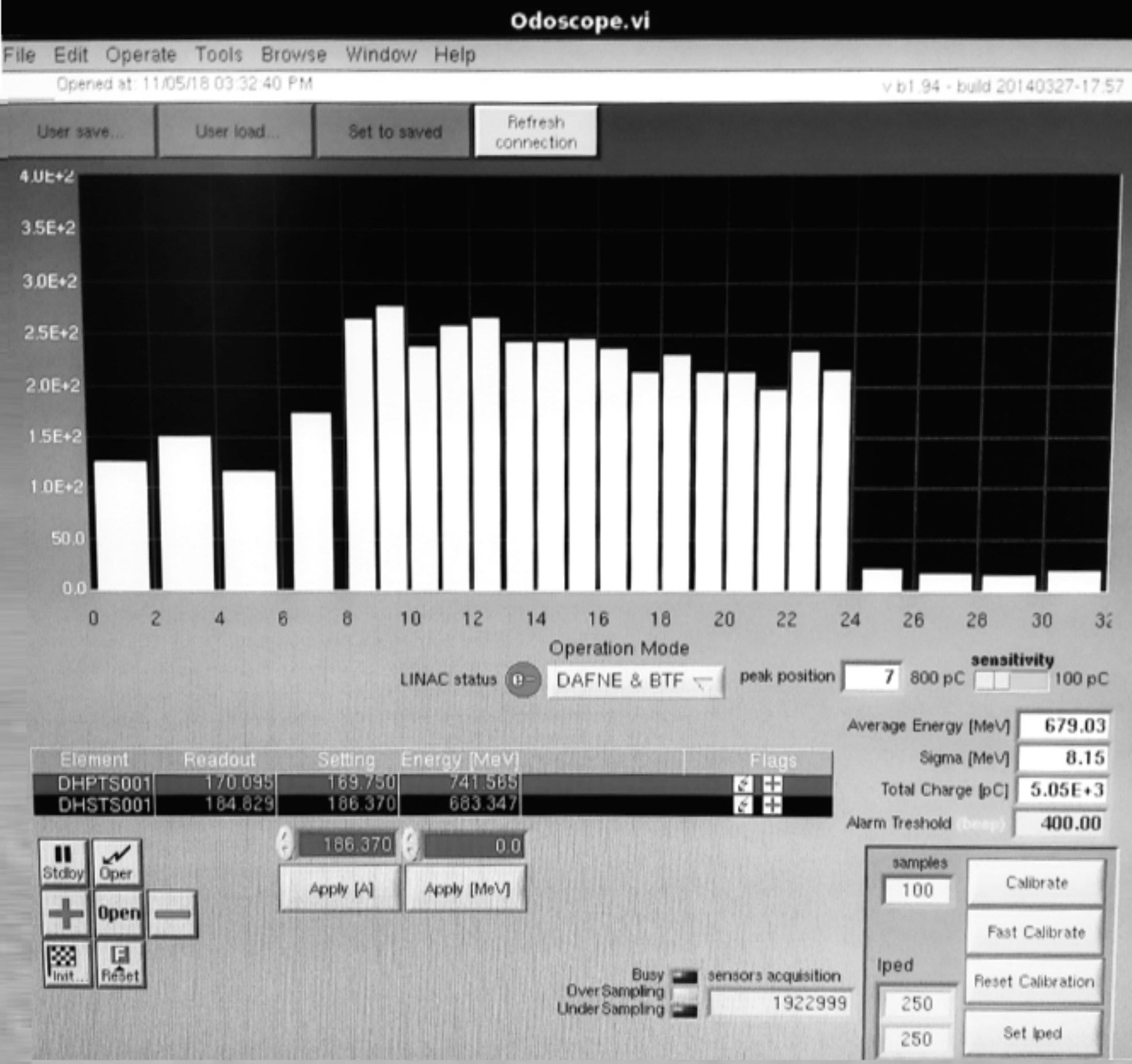}
 \caption{
      Beam spread of 200~ns long beam pulses, measured by the LINAC spectrometer.
    \label{fig8} }
\end{figure}

The effect of accelerating voltage shape is clearly changing the delay of the gun pulse with respect to the optimal set for the four modulators: shifting further in advance or delaying the beam passage, results in a loss of intensity either at the head or at the tail of the macro-bunch, as shown in Fig.~\ref{fig7}. Moreover, a different time along the beam pulse will correspond to a different energy, due to the shape of the accelerating voltage. As a result, the beam will have a much larger spread: from a typical value of $<0.5$\% with 10~ns long pulses, a spread larger by approximately one order of magnitude ($\sim$6\%) has been achieved at 200~ns, as measured by the LINAC spectrometer (see Fig.~\ref{fig8}). 

Looking at the energy distribution measured by the spectrometer was also possible to verify the time-energy correlation: the higher part of the spectrum corresponds to the peak of the accelerating voltage, i.e. the tail of the long macro-bunch, while lower energy particles populate the head. 

\subsection{Positron production}
The positron beam delivered to PADME can be either ``primary'' or ``secondary''. 
The first case refers to positrons produced by $\sim$200~MeV electrons hitting a production target (positron converter, two radiation lengths W-Re) placed after the first five 3~m accelerating sections of the LINAC. The produced positrons are focussed by a system of solenoids and then accelerated by the remaining ten accelerating sections of the LINAC, up to 550~MeV. This is the standard way of producing the high-charge, 510~MeV, 10~ns long positron pulses for injection into the DA$\Phi$NE collider.

\begin{figure}[htbp]
\centering
\includegraphics[width=\textwidth]{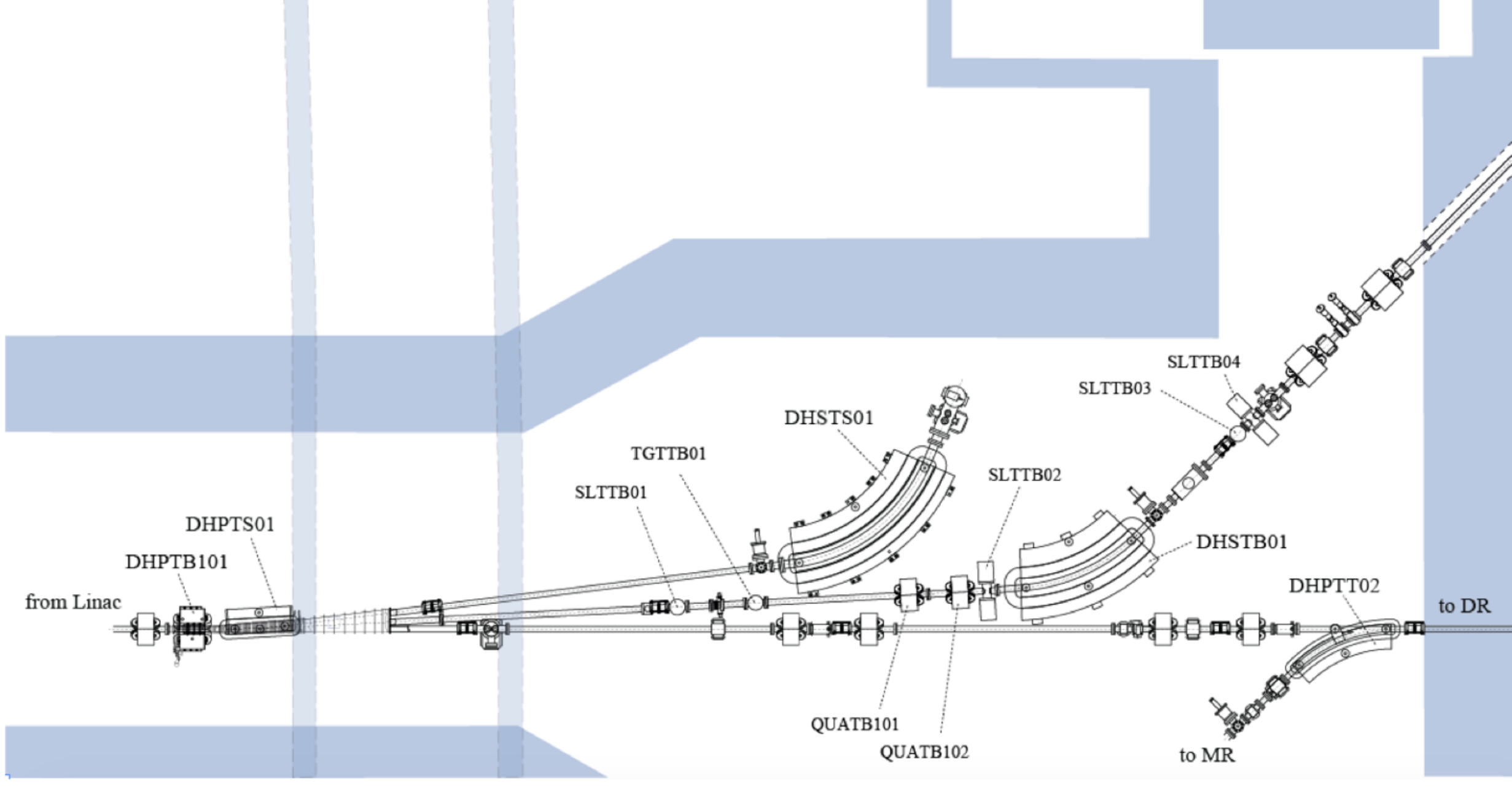}
 \caption{
      Layout of the BTF transfer-line with the Cu target (TGTTB01), the energy-selecting dipole (DHSTB01) and horizontal (SLTTB02/SLTTB04) and vertical (SLTTB01/SLTTB03) collimators.
    \label{fig9} }
\end{figure}

In the ``secondary'' beam configuration, all the sections are used to accelerate electrons from the gun close to the maximum energy, e.g. $\sim$700~MeV, which are then driven onto the beam-test line. At the beginning of the BTF transfer-line (shown in Fig.~\ref{fig9}) they hit a two radiation lengths Cu target (TGTTB01), producing a wide spectrum of secondary particles. Positrons of the required momentum are then selected by a magnetic system made by a sector dipole (DHSTB01) and a horizontal collimator (SLTTB04) which, together with the upstream horizontal collimator (SLTTB02) also defines the momentum spread~\cite{ref15}. In this case, the intensity of the positron beam is strongly dependent on the selected energy.

\section{Extending the beam pulse}
In the following paragraphs, different techniques for further extending the LINAC positron pulse beyond $\approx200$~ns are briefly described.

Two simple principles have been taken into account in considering the different options: 
\begin{itemize}
\item the basic requirement for the $X$(17~MeV) search is to accelerate positrons at energy of $\leq$300~MeV, i.e. at considerably lower than the maximum achievable by the LINAC. In particular, this is approximately a factor two lower than the maximum energy, which corresponds roughly to the increase of gradient due to the SLED.
\item Another important point is to keep the possibility of reverting to the standard operation of the LINAC for the production of 510 MeV, 10 ns long, high current positron pulses, for the injection into the ring. This is a very important requirement as long as one wants to keep the possibility of running DA$\Phi$NE as electron-positron collider. 
\end{itemize}
The proposed methods have different levels of complexity and impact on the LINAC hardware, and not all of them have been routinely used in similar accelerator facilities, and thus would require different R\&D and testing.

\subsection{Extending the beam pulse: LLRF modulation}
The most efficient approach for getting a flat-top output from the SLED output is to modulate the phase of the input RF, instead of making just one reversal. This can be achieved by continuous phase modulation up to the $180^\circ$ shift~\cite{ref16}. The modulation can be as simple as linear phase increase with an optimized slope~\cite{ref17}, and generally can implemented using a digital LLRF system. More sophisticated approaches foresee the modulation both of the input RF phase and amplitude~\cite{ref18}.

Implementing this solution for the DA$\Phi$NE LINAC could allow getting a flat-top, and thus a very low-energy spread beam (in principle 0.5\% or smaller, as achieved with very short pulses), over all the SLED time ($t_2-t_1$)=800~ns, with a somewhat intermediate energy gain between 1 (no SLED) and 1.6 (full compression), i.e. in the range of 400~MeV for positrons. 
However, this requires the implementation of a new, fully digital system in the LLRF.

\subsection{Extending the beam pulse: manipulating the phase inversion}
\subsubsection{Changing phase inversion time}
The effect of changing the delay of the PSK signal for triggering the phase inversion has been studied on a SLED practically identical to the DA$\Phi$NE one at the BEPC injector~\cite{ref19}: by anticipating the inversion time $t_1$ without changing the closing time $t_2$ (referring to Fig.~\ref{fig4}), results in a slightly lower peak and of course a longer compressed RF time. Even though the slope of the falling edge of the compressed RF pulse should not be affected, the acceleration of longer beam pulses should be possible. 

\subsubsection{Multiple phase inversions}
As described in Ref.~\cite{ref13} beam pulses up to $\sim$150~ns, with good energy spread, were produced at high-charges in the SLAC LINAC, thanks to the beam-loading compensation also used for the PADME Run-1 (as briefly discussed in Sec.~\ref{run1}).

In order to further extend the beam pulses also at lower beam charges, a new technique was implemented at SLAC, by generating a notch in the SLED RF output. This was realized by generating a pair of additional phase switches after the normal 180$^\circ$ one. The notch in that case was generated 280~ns after the first phase flip at the beginning of the SLED pulse (at time $t_1$) and lasted for 400~ns. In this way the RF energy is stored again in the SLED cavities, producing a higher RF output towards the end ($t_2$). In this configuration, up to 460~ns long electron pulses with 0.56~mA current were produced with reduced energy spread. 

The maximum energy was only slightly reduced (by $\sim$7\%) since the additional phase inversions were implemented only on 20\% of the SLAC LINAC (6 sectors out of 30). Scaling that result to the DA$\Phi$NE LINAC, one can expect to get up to 500~ns long positron pulses at a maximum energy of $\sim$350~MeV, assuming a similar implementation of the RF notch on all four RF stations.
The implementation of the two additional phase reversals requires sending two additional trigger signals, with the correct timing, to the bi-phase modulator generating the PSK in the new LLRF driver~\cite{ref20}.  

\subsubsection{No phase inversion}
The role of the phase inversion is to enhance the effect of the discharge of the RF energy stored in the two SLED cavities, by adding (with the same phase) the RF wave from the klystron transmitted to the accelerator through the 3 dB hybrid to the wave re-radiated by the filled cavities, which gets a 180$^\circ$ shift in the reflection.

However, if one removes the phase jump, and just lets the SLED cavities fill during from $t_0$ to $t_2$, and then re-radiate the RF wave immediately after the klystron is switched off, the resulting output power will still have a peak higher than the incoming one, and will just decay exponentially with the same time constant of filling $\tau_F=2Q_L/\omega$.

\subsection{Extending the beam pulse: SLED detuning}
An obvious method for getting a long and flat RF pulse to the accelerating structures is to completely by-pass the SLED. There are several methods, which differ mainly in the possibility and ease in switching from the uncompressed operation back to the fully compressed operation for short pulses. This is necessary if one wants to keep the possibility of reverting in a practical and reliable way to the acceleration of 10~ns long, $\sim$1~nC charge, 510~MeV electron and positron beams for injections into the DA$\Phi$NE damping ring.

\subsubsection{Detuning needle}
The original design of the SLED device already includes a quick and controllable way of controlling the tuning of the cavities: a Tungsten, thin needle, that can be pushed inside the cylindrical cavity at a well-defined angle, such that when fully inserted the tip reaches a a circle of maximum azimuthal electric field. This causes a shift of the frequency by approximately 100 times the cavity bandwidth, corresponding to a negligible power.
This is the standard tool used for fine adjusting the frequency of the SLED cavities, which is generally done by minimizing the reflected RF wave at low power. 

Unfortunately, the complete insertion and removal of the needles in some of the $4\times2$ cavities has failed during the lifetime of the LINAC. Fixing this problem would imply a major hardware intervention, not only resource and time-consuming, but which can also result in the impossibility of tuning the device back to the resonance condition. As much as one wants to keep the possibility of going back to the conditions of injection into the ring, this is a serious issue.

A possible solution, which of course requires time and money, it to realize new SLED’s for replacing the ones with failing tuning device.

\subsubsection{Temperature shift}
It has been reported that the cavities tuning can be controlled by regulating the temperature of the water circuit of the SLED devices: in particular in the Australian Synchrotron injector (S-band, 100~MeV LINAC using two pulsed 37~MW klystrons) lowering the SLED operating temperature from 40$^\circ$C to 20$^\circ$C is sufficient to move the frequency of the cavities off enough to let the RF pulse bypass the input~\cite{ref21}. . 

The parameters of this SLED device~\cite{ref22} are close to those of the DA$\Phi$NE LINAC one: $\beta=3.5$ and $Q_0=10^5$, operating temperature $40^\circ$C. It is used to feed S-band cavities accelerating electron macro-pulses of 140~ns length.
The temperature coefficient is $\sim$45~kHz/$^\circ$C (as measured at SLAC and IHEP~\cite{ref18}), which has to be compared to the bandwidth $\Delta f=f/Q_L\simeq170$~kHz. At this level, a significant effect is also expected from relative humidity variations (a temperature-dependent coefficient is expected). 

As a simple and practical solution, one can consider feeding the SLED water circuit at the temperature of the primary cooling circuit, $\sim$18$^\circ$C, instead of the standard operating temperature of SLED and accelerating structures of $\sim$43$^\circ$C. This would give a frequency change of 1125~kHz, that should be sufficient for shifting the cavities off-resonance, since it corresponds to $\approx$7 times the bandwidth, i.e. a power few \% of the resonance value. 

Alternatively, one can consider going to higher operating temperature. The practical advantage, in this case, would be the possibility of adding a dedicate heat exchanger for setting the SLED temperature higher than the LINAC secondary water circuit set-point.

\subsubsection{Cavities unbalancing}
\begin{figure}[htbp]
\centering
\includegraphics[width=0.7\textwidth]{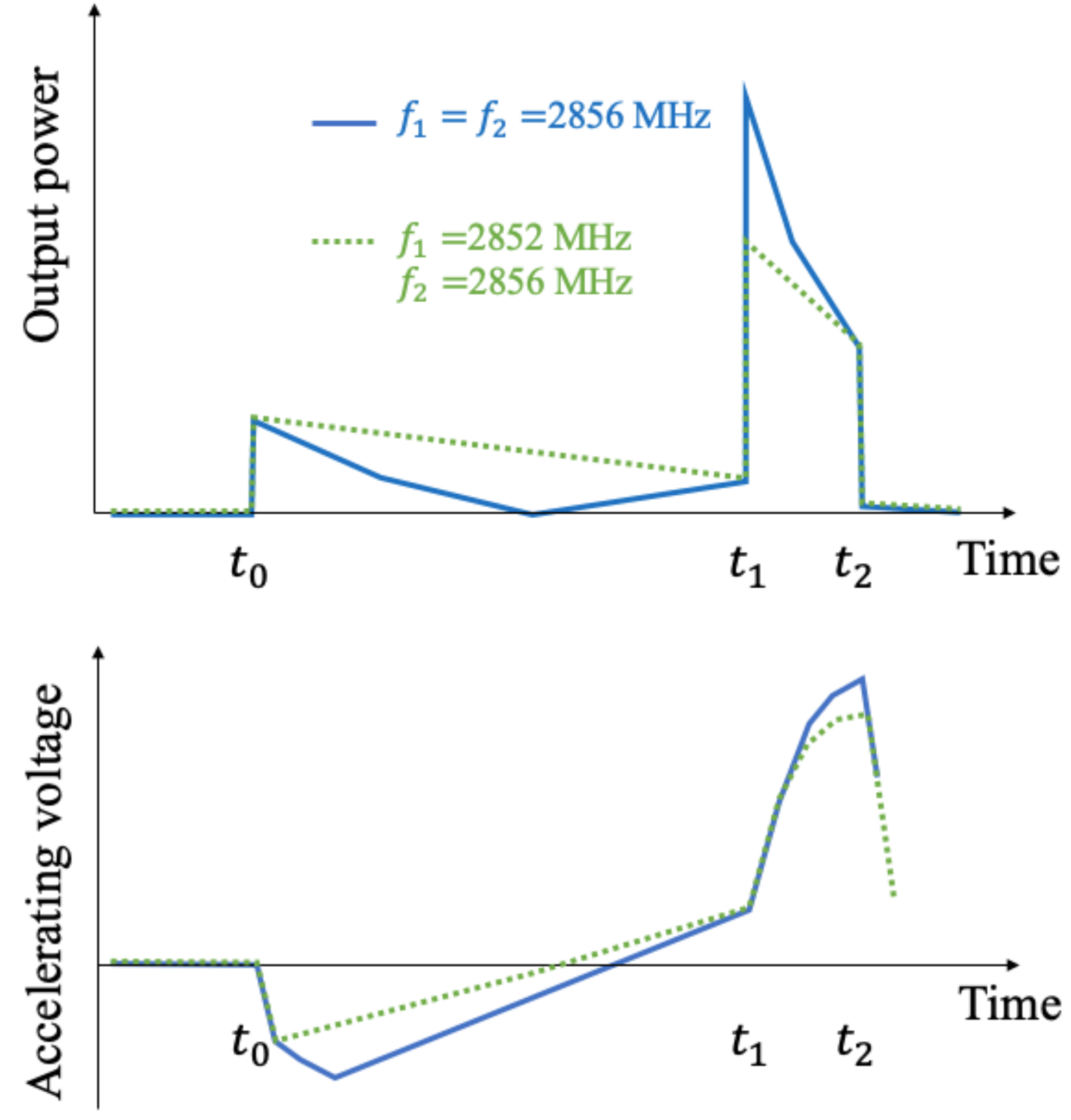}
 \caption{
  Effect (qualitative) of unbalancing the two SLED cavities in terms of SLED output power (top, adapted from Ref.~\cite{ref18}) and accelerating voltage (bottom).
    \label{fig10} }
\end{figure}

The effect of detuning of only one of the two SLED cavities, with respect to the resonance frequency, has been studied in Ref.~\cite{ref18}: the effect on the compressed RF output is to reduce the peak height, i.e. the energy gain, and thus to reduce the slope of the falling edge, as shown in Fig.~\ref{fig10}. This should result in smoother accelerating voltage shape, then allowing the acceleration of longer beam pulses (given the maximum allowable energy spread in the LINAC).

A significant fraction of the RF power will be reflected, but the SLED will still function, even though with a lower energy gain.

\subsubsection{Beam intensity with detuned SLED}
Increasing the duration of the pulse on the LINAC gun will extract a linearly increasing charge, i.e. the current from the cathode and the extraction system is constant, up to several tens of ns~\cite{ref11}. Further increasing the pulse width produces a beam charge out of the maximum range of the current monitors (10~nC) even reducing the gun filament current. Even with short positron pulses for DA$\Phi$NE injections (10~ns), charges exceeding 0.5 nC have been routinely achieved (corresponding to 1.5$\times10^{11}$ particles/s at the maximum repetition rate of 49 pulses/s). 

Indeed, when running with long pulses during the PADME run, the charge extracted from the gun cathode was strongly reduced, both decreasing the applied voltage and by adjusting the grid counter-potential. Running with a much-reduced charge already in the first part of the LINAC required the optimization of the LINAC magnetic optics, starting from the focussing solenoids around the first sections (``Helmholtz'' coils), including the quadruples around the LINAC sections and downstream in the transfer-line. 

In any case, the main limitation on the beam intensity will be given by the radio-protection rules for the BTF hall, allowing a maximum average flux of 3.125$\times10^{10}$ particles/s at 510~MeV. Translating this figure in terms of beam power, this corresponds to 2.55~W, i.e. a maximum flux of 0.53$\times10^{11}$ particles/s at~300 MeV.

\section{Conclusions}
Searching with the PADME experiment at the BTF the hypothetical 17 MeV/$c^2$ dark boson that could explain the 8Be anomaly in thick-target, positron on target annihilations requires to run the DA$\Phi$NE LINAC close to the resonance, $E_+\leq 300$ MeV, with the longest possible beam pulse in order to reach the highest possible luminosity, while keeping the detector pile-up and over-veto under control.

Secondary (primary) positron beam pulses up to $\sim$550 (490)~MeV energy, with length $\sim$200~ns were successfully produced and accelerated during the PADME first data-taking period, using the standard RF configuration of the LINAC with SLED compression, originally designed for the acceleration of short ($\sim10$~ns), high-charge electron and positron beams for the injection in the DA$\Phi$NE collider.

Several methods for further extending the positron beam pulses have been proposed, with two simple inspiring principles. The first is that since the basic requirement for the $X$(17~MeV) search is to accelerate positrons at energy of $\leq$300~MeV, it is possible to get a flat accelerating voltage by reducing (or even eliminating) the effect of SLED on the klystron power, considering that an energy in the required range can be achieved with lower or even no gain from the RF compression. 

The other aspect to be taken into account is to keep the possibility of reverting to the standard high-energy, short-pulses LINAC operation in a reliable, easy and practical way.

The most straightforward method would be making the RF wave completely by-pass the SLED, either by inserting the de-tuning needle into the resonant cavities or by bringing them completely off-resonance, e.g. by changing drastically the operating temperature. Such a method would produce a no-gain, completely flat, 4.5~$\mu$s long, RF power, that -- taking into account the filling time of the accelerating structures -- should produce a $\sim$300~MeV positron beam with macro-bunch length of the order of 3~$\mu$s (once such a long electron pulse is produced at the gun). If the insertion of the tuning needle has been problematic in the past (and even impossible in one out of the four devices), on the other hand the temperature-driven detuning has never been tested in the DA$\Phi$NE LINAC.

Another reliable approach would be keeping the SLED devices tuned to the resonance, but then modulating the low-level RF signal. Even in the simplest form of a linear-slope phase increase, as developed in the ELETTRA injector LINAC, a completely flat, $\sim$800~ns long RF pulse would be produced, thus allowing the acceleration of a similar length positron bunch, with a maximum energy slightly lower than the maximum, but well above the required 300 MeV value. However, this requires a new (digital) implementation of the phase ramp in the LLRF system.

Other solutions would allow getting pulses in an intermediate range, from 200~ns to 800~ns, essentially by smoothing the RF compressed pulse or extending the small flat-top. This can be achieved using different techniques, such as: adding a pair of 180$^\circ$ phase jumps, removing the phase inversion or changing its timing, or detuning only one SLED cavity. None of these methods has been previously tested in the DA$\Phi$NE LINAC, but they all require small modifications to the RF system or smaller impact operations (like detuning of only one cavity).

In order to assess the achievable positron beam length, energy and quality at least for the options requiring less important modifications to the LINAC, a test campaign would be extremely useful. The essential program of such a testing should include the measurement of the SLED output power without PSK or with different PSK timing, and by changing the operating temperature. As final consideration, the availability of a full RF test-station, including modulator, klystron and SLED would be greatly useful for performing this kind of studies in parallel to the standard LINAC operations.

\end{document}